\documentclass[twocolumn,prl]{revtex4}

\usepackage{amsmath,amsthm,amsfonts,amscd,amssymb} 
\usepackage{eucal}
\usepackage{braket}
\usepackage{slashed}
\usepackage{hyperref}
\usepackage{graphicx}
\usepackage{bm}

\usepackage{amsmath,amsthm,amsfonts,amscd,amssymb} 
\usepackage{eucal}
\usepackage{braket}
\usepackage{slashed}
\usepackage{hyperref}
\usepackage{graphicx}
\usepackage{bm}

\newcommand{\Oi}{\mathcal{O}}

\newcommand{\be}{\begin{equation}}
\newcommand{\ee}{\end{equation}} 
\newcommand{\mb}{\mathbf}

\newcommand{\bear}{\begin{eqnarray}}
\newcommand{\eear}{\end{eqnarray}}

\begin{document}

\title{Infrared quantum information}

  \author{Daniel Carney, Laurent Chaurette, Dominik Neuenfeld, Gordon Walter Semenoff}
     \affiliation{
	  Department of Physics and Astronomy
	  University of British Columbia\\ 6224 Agricultural Road, 
	  Vancouver, BC V6T 1Z1 Canada }

\begin{abstract}
We discuss information-theoretic properties of low-energy photons and gravitons in the $S$-matrix. Given an incoming $n$-particle momentum eigenstate, we demonstrate that unobserved soft photons decohere nearly all outgoing momentum superpositions of charged particles, while the universality of gravity implies that soft gravitons decohere nearly all outgoing momentum superpositions of \emph{all} the hard particles. Using this decoherence, we compute the entanglement entropy of the soft bosons and show that it is infrared-finite when the leading divergences are re-summed \`{a} la Bloch and Nordsieck.
\end{abstract}

\maketitle

The massless nature of photons and gravitons leads to an infrared catastrophe, in which the $S$-matrix becomes ill-defined due to divergences coming from low-energy virtual bosons. The usual solution to this problem, originally given by Bloch and Nordsieck in electrodynamics \cite{Bloch:1937pw} and extended to gravity by Weinberg \cite{Weinberg:1965nx}, is to argue that an infinite number of low-energy bosons are radiated away during a scattering event; this leads to divergences which cancels the divergences from the virtual states, and physical predictions in terms of infrared-finite inclusive transition probabilities.

In this letter, we study quantum information-theoretic aspects of this proposal. Since each photon and graviton has two polarization states and three momentum degrees of freedom, one might suspect that the low-energy radiation produced during scattering could carry a huge amount of information. Here we demonstrate that, according to the methodology of \cite{Bloch:1937pw,suura,Weinberg:1965nx}, if the initial state is an incoming $n$-particle momentum eigenstate, the ``soft'' bosonic divergences can lead to complete decoherence of the momentum state of the outgoing ``hard'' particles. This decoherence is avoided only for superpositions of pairs of outgoing states for which an infinite set of angle-dependent currents match, see eq. \eqref{currents}. In simple examples like QED, this will be enough to get complete decoherence of all momentum superpositions. In less simple cases, one is still left with an extremely sparse density matrix dominated by its diagonal elements.

Having traced the radiation in this fashion, we obtain an infrared-finite, mixed reduced density matrix for the hard particles. In the simple cases when we get a completely diagonal matrix, we compute the entanglement entropy carried by the soft gauge bosons. The answer is finite and scales like the logarithm of the energy resolution $E$ of a hypothetical soft boson detector.

While the tracing out of the soft radiation can be viewed as a physical statement about the energy resolution of a real detector, in this formalism, the trace is also forced on us by mathematical consistency: it is the only way to get well-defined transition probabilities from the infrared-divergent $S$-matrix. There is an alternative approach to the infrared catastrophe, in which one constructs an IR-finite $S$-matrix of transition amplitudes between ``dressed'' matter states.\cite{Chung:1965zza,kibble,Kulish:1970ut,Ware:2013zja} In such an approach, there are no divergences and so one is not forced to trace over any soft radiation. Whether the two formalisms lead to the same physical picture is an interesting question, and we leave a detailed comparison to future work.

Recently, the infrared structure of gauge theories has become a topic of much interest due to the proposal that soft radiation may encode information about the history of formation of a black hole.\cite{Hawking:2016msc,Hawking:2016sgy,Mirbabayi:2016axw} We hope that our work can make this discussion more quantitatively grounded; we comment on black holes at the end of this letter. More generally, it is of interest to understand the information-theoretic nature of the infrared sector of quantum field theories, and our paper is intended to make some first steps in this direction.

\emph{Decoherence of the hard particles.} Fix a single-particle energy resolution $E$. We define soft bosons as those with energy less than $E$, and hard particles as anything else. Consider an incoming state $\ket{\alpha}_{in}$ consisting of hard particles, charged or otherwise, of definite momenta.\footnote{Our field theory conventions follow \cite{Weinberg:1995mt}. Labels like $\alpha,\beta,b$ mean a list of free-particle quantum numbers, e.g. $\ket{\alpha}_{in} = \ket{\mb{p}_1 \sigma_1, \ldots}_{in}$ listing momenta and spin of the incoming particles.} The $S$-matrix evolves this into a coherent superposition of states with hard particles $\beta$ and soft bosons $b = \gamma, h$ (photons $\gamma$ and gravitons $h$),
\be
\ket{\alpha}_{in} = \sum_{\beta b} S_{\beta b,\alpha} \ket{\beta b}_{out}.
\ee
Hereafter we drop the subscript on kets, which will always be out-states. Tracing out the bosons $\ket{b}$, the reduced density matrix for the outgoing hard particles is
\be
\label{DMelement}
\rho = \sum_{\beta \beta'  b}  S_{\beta b,\alpha} S^*_{\beta' b,\alpha} \ket{\beta}\bra{\beta'}.
\ee
Using the usual soft factorization theorems \cite{Low:1958sn,suura,Weinberg:1965nx}, we can write the amplitudes in terms of the amplitudes for $\alpha \to \beta$ multiplied by soft factors, one for each boson:
\be
\label{softsumresult}
S_{\beta b,\alpha} = S_{\beta,\alpha} F_{\beta,\alpha}(\gamma) G_{\beta,\alpha}(h),
\ee
where the soft factors $F,G$ are
\begin{align}
\begin{split}
F_{\beta,\alpha}(\gamma) & = \sum_{n \in \alpha,\beta} \sum_{\pm} \prod_{i \in \gamma} \frac{e_n \eta_n }{(2\pi)^{3/2} |\mb{k}_i|^{1/2}} \frac{p_n^{\mu} \epsilon^*_{\mu,\pm}(\mb{k}_i)}{p_n \cdot k_i - i \eta_n \epsilon} \\
G_{\beta,\alpha}(h) & = \sum_{n \in \alpha,\beta} \sum_{\pm} \prod_{i \in h} \frac{M_{p}^{-1} \eta_n }{(2\pi)^{3/2} |\mb{k}_i|^{1/2}} \frac{p_n^{\mu} p_n^{\nu} \epsilon^*_{\mu\nu,\pm}(\mb{k}_i)}{p_n \cdot k_i - i \eta_n \epsilon}. 
\end{split}
\end{align}
Here the index $n$ runs over all the incoming and outgoing hard particles, $i$ runs over the outgoing soft bosons; $\eta_n = -1$ for an incoming and $+1$ for an outgoing hard particle. The $e_n$ are electric charges and $M_p = (8\pi G_N)^{-1/2}$ is the Planck mass, and the $\epsilon$'s are polarization vectors or tensors for outgoing soft photons and gravitons, respectively. By an argument identical to the one employed by Weinberg \cite{Weinberg:1965nx}, and assuming we can neglect the total lost energy $E_T$ compared to the energy of the hard particles, we can use this factorization to perform the sum over soft bosons in \eqref{DMelement}, and we find that
\begin{align}
\begin{split}
\label{realdivanswer}
\sum_{b} S_{\beta b,\alpha} S^*_{\beta' b,\alpha} & = S_{\beta,\alpha} S^*_{\beta',\alpha} \left( \frac{E}{\lambda} \right)^{\tilde{A}_{\beta\beta',\alpha}} \left( \frac{E}{\lambda} \right)^{\tilde{B}_{\beta\beta',\alpha}} \\
& \times f\left(\frac{E}{E_T},\tilde{A}_{\beta\beta',\alpha}\right) f\left(\frac{E}{E_T},\tilde{B}_{\beta\beta',\alpha}\right).
\end{split}
\end{align}
Here $\lambda \ll E$ is an infrared regulator used to cut off momentum integrals which we will send to zero later; one can think of $\lambda$ as a mass for the photon and graviton. The exponents are
\begin{align}
\label{tildeABdef}
\begin{split}
\tilde{A}_{\beta\beta',\alpha} & = -\sum_{\substack{n \in \alpha,\beta \\ n' \in {\alpha,\beta'}}} \frac{e_n e_{n'} \eta_n \eta_{n'}}{8\pi^2} \beta_{nn'}^{-1} \ln \left[ \frac{1+\beta_{nn'}}{1-\beta_{nn'}} \right] \\
\tilde{B}_{\beta\beta',\alpha} & = \sum_{\substack{n \in \alpha,\beta \\ n' \in {\alpha,\beta'}}} \frac{m_n m_{n'} \eta_n \eta_{n'}}{16 \pi^2 M_p^2} \frac{1+\beta_{nn'}^2}{\beta_{nn'} \sqrt{1 - \beta_{nn'}^2}} \ln \left[ \frac{1+\beta_{nn'}}{1-\beta_{nn'}} \right],
\end{split}
\end{align}
and $f$ is a complicated function which can be found in \cite{Weinberg:1995mt}; for $E/E_T = \Oi(1)$ and for small $A$, $f$ may be approximated as $f(1,A) \approx  1 - \pi^2 A^2/12 + \Oi(A^4)$. In these formulas, $\beta_{nn'}$ is the relative velocity between particles $n$ and $n'$,
\begin{equation*}
\beta_{nn'} = \sqrt{1 - \frac{m_n^2 m_{n'}^2}{(p_n \cdot p_{n'})^2}},
\end{equation*}
For future use, we note that $0 \leq \beta \leq 1$, and both of the dimensionless functions of $\beta$ appearing in \eqref{tildeABdef} run over $[2,\infty)$ as $\beta$ runs from $0$ to $1$. We have $\beta_{nm} = 0$ if and only if $p_n = p_m$.

The divergences as $\lambda \to 0$ in \eqref{realdivanswer} come from summing over an infinite number of radiated, on-shell bosons. There are also infrared divergences inherent to the transition amplitude $S_{\beta,\alpha}$ itself coming from virtual bosons. Again following Weinberg, we can add these divergences up, and we have that
\be
\label{virtualdivergences}
S_{\beta,\alpha} = S_{\beta,\alpha}^{\Lambda} \left( \frac{\lambda}{\Lambda} \right)^{A_{\beta,\alpha}/2} \left( \frac{\lambda}{\Lambda} \right)^{B_{\beta,\alpha}/2},
\ee
where now $S_{\beta,\alpha}^{\Lambda}$ means the amplitude computed using only virtual bosons of energy above $\Lambda$, and
\begin{align}
\label{ABdef}
\begin{split}
A_{\beta,\alpha} & = -\sum_{n,m \in \alpha,\beta} \frac{e_n e_m \eta_n \eta_m}{8\pi^2} \beta_{nm}^{-1} \ln \left[ \frac{1+\beta_{nm}}{1-\beta_{nm}} \right] \\
B_{\beta,\alpha} & =  \sum_{n,m \in \alpha,\beta}  \frac{m_n m_m \eta_n \eta_m}{16 \pi^2 M_p^2} \frac{1+\beta_{nm}^2}{\beta_{nm} \sqrt{1 - \beta_{nm}^2}} \ln \left[ \frac{1+\beta_{nm}}{1-\beta_{nm}} \right].
\end{split}
\end{align}
An infrared-divergent ``Coulomb'' phase is suppressed in \eqref{virtualdivergences}. We will see shortly that this phase cancels out of all the relevant density matrix elements. 

Putting the above results together, we find that the reduced density matrix coefficient for $\ket{\beta} \bra{\beta'}$ is given by
\begin{align}
\begin{split}
\label{DMelementgeneral}
\rho_{\beta\beta'} & = S^{\Lambda}_{\beta,\alpha} S^{\Lambda*}_{\beta',\alpha} \left( \frac{E}{\lambda} \right)^{\tilde{A}_{\alpha,\beta\beta'}} \left( \frac{\lambda}{\Lambda} \right)^{A_{\beta,\alpha}/2+A_{\beta',\alpha}/2} \\
& \times \left( \frac{E}{\lambda} \right)^{\tilde{B}_{\alpha,\beta\beta'}} \left( \frac{\lambda}{\Lambda} \right)^{B_{\beta,\alpha}/2+B_{\beta',\alpha}/2}  f(\tilde{A}_{\beta\beta',\alpha}) f(\tilde{B}_{\beta\beta',\alpha}).
\end{split}
\end{align} 
The question is how this behaves in the limit that the infrared regulator $\lambda \to 0$. The coefficient scales as $\lambda^{\Delta A+\Delta B}$, where
\begin{align}
\label{differentialexps}
\begin{split}
\Delta A_{\beta\beta',\alpha} & = \frac{A_{\beta,\alpha}}{2} + \frac{A_{\beta',\alpha}}{2} - \tilde{A}_{\beta\beta',\alpha} \\
\Delta B_{\beta\beta',\alpha} & = \frac{B_{\beta,\alpha}}{2} + \frac{B_{\beta',\alpha}}{2} - \tilde{B}_{\beta\beta',\alpha}.
\end{split}
\end{align}
In the appendix, we prove that both of these exponents are positive-definite, $\Delta A_{\beta\beta',\alpha} \geq 0$ and $\Delta B_{\beta\beta',\alpha} \geq 0$. The density matrix components \eqref{DMelementgeneral} which survive as the regulator $\lambda \to 0$ are those for which $\Delta A = \Delta B = 0$; all other density matrix elements will vanish.

To give necessary and sufficient conditions for $\Delta A = \Delta B = 0$, we define two currents for each spatial velocity vector $\mb{v}$. We assume for simplicity that only massive particles carry electric charge. For massive particles, there are electromagnetic and gravitational currents defined as
\begin{align}
\begin{split}
\label{currents}
j^{EM}_{\mb{v}} & = \sum_i e^i a_{\mb{p}_i(\mb{v})}^{i\dagger} a_{\mb{p}_i(\mb{v})}^i \\
j^{GR}_{\mb{v}} & = \sum_i E_i(\mb{v}) a_{\mb{p}_i(\mb{v})}^{i\dagger} a_{\mb{p}_i(\mb{v})}^i.
\end{split}
\end{align}
Here $i$ labels particle species, $e^i$ their charges and $m^i$ their masses; the kinematic quantities $\mb{p}_i(\mb{v}) = m_i \mb{v}/\sqrt{1-\mb{v}^2}$ and $E_i(\mb{v}) = m_i/\sqrt{1-\mb{v}^2}$ are the momentum and energy of species $i$ when it has velocity $\mb{v}$. For lightlike particles we have to separately define the gravitational current, since a velocity and species does not uniquely determine a momentum:
\be
\label{masslesscurrents}
j^{GR,m=0}_{\mb{v}} = \sum_i \int_0^{\infty} d\omega \ \omega a_{\omega \mb{v}}^{i\dagger} a_{\omega \mb{v}}^i.
\ee
Momentum eigenstates of any number of particles are obviously eigenstates of these currents and we denote their eigenvalues $j_{\mb{v}} \ket{\alpha} = j_{\mb{v}}(\alpha) \ket{\alpha}$.

The photonic exponent $\Delta A_{\beta\beta',\alpha}$ is zero if and only if the charged currents in $\beta$ are the same as those in $\beta'$; the gravitational exponent $\Delta B_{\beta\beta',\alpha}$ is zero if and only if $\emph{all}$ the hard gravitational currents in $\beta$ are the same as those in $\beta'$. This is demonstrated in detail in the appendix. For any such pair of outgoing states $\ket{\beta},\ket{\beta'}$, \eqref{DMelementgeneral} becomes independent of the IR regulator $\lambda$ and is thus finite as $\lambda \to 0$,
\be
\label{DManswer}
\rho_{\beta\beta'} = S^{\Lambda *}_{\beta'\alpha}  S^{\Lambda}_{\beta\alpha} \mathcal{F}_{\beta\alpha} \left( E, E_T, \Lambda \right),
\ee
where
\be
\label{calF}
\mathcal{F}_{\beta\alpha} =  f\left(\frac{E}{E_T},A_{\beta\alpha} \right) f\left(\frac{E}{E_T},B_{\beta\alpha} \right) \left( \frac{E}{\Lambda} \right)^{A_{\beta\alpha} + B_{\beta\alpha}}.
\ee
This is the case in particular for diagonal density matrix elements $\beta = \beta'$, for which we obtain the standard transition probabilities
\be
\label{DManswer2}
\rho_{\beta\beta} = \left| S^{\Lambda}_{\beta\alpha} \right|^2 \mathcal{F}_{\beta\alpha} \left( E, E_T, \Lambda \right).
\ee
On the other hand, if there is even a single $\mb{v}$ for which one of the currents \eqref{currents} or \eqref{masslesscurrents} does not have the same eigenvalue in $\ket{\beta}$ and $\ket{\beta'}$, then the density matrix coefficient decays as $\lambda^{\Delta A + \Delta B} \to 0$ as the regulator $\lambda \to 0$. We see that the unobserved soft bosons have almost completely decohered the momentum state of the hard particles. Only a very sparse subset of superpositions in which all the $j_{\mb{v}}(\beta) = j_{\mb{v}}(\beta')$ survive.

\emph{Examples}. To get a feel for the results presented in the previous section, we consider a few examples. First, consider any scattering with a single incoming and outgoing charged particle, like potential or single Compton scattering. Let the incoming momentum be $\alpha = p$ and the outgoing momenta of the two branches $\beta = q, \beta' = q'$. We have either directly from the definition \eqref{differentialexps} or the theorem \eqref{DeltaAstatement} that
\be
\Delta A_{qq',p} = -\frac{e^2}{8\pi^2} \left[ 2 - \gamma_{qq'} \right],
\ee
where $\gamma_{qq'} = \beta_{qq'}^{-1} \ln (1+\beta_{qq'})/(1-\beta_{qq'})$. This $\Delta A$ is easily seen to equal zero if and only if $q=q'$. Thus other than the spin degree of freedom, the resulting density matrix for the charge is exactly diagonal in momentum space. 

To see an example where the current-matching condition is non-trivially fulfilled, consider a theory with two charged particle species of charge $e$ and $e/2$ and the same mass. Then we can get an outgoing superposition of a state $\beta = (e,q)$ and one with two half-charges $\beta' = (e/2,q'_1) + (e/2,q'_2)$. The differential exponent for such a superposition is
\be
\Delta A_{\beta\beta',p} = -\frac{e^2}{8\pi^2} \left[ 3 + \frac{1}{2} \gamma_{q_1 q_2} - \gamma_{q q_1} - \gamma_{q q_2} \right],
\ee
which is zero if $q = q_1 = q_2$. In other words, the currents \eqref{currents} cannot distinguish between a full charge of momentum $q$ and two half-charges of the same momentum.

\emph{Entropy of the soft bosons.} We have seen that the reduced density matrix for the outgoing hard particles is very nearly diagonal in the momentum basis. In a simple example like a theory with various scalar fields $\phi_i$ of different, non-zero masses $m_i$, the soft graviton emission causes \emph{complete} decoherence into a diagonal momentum-space reduced density matrix for the hard particles. More generally, we may have a sparse set of superpositions, and in any case spin and other internal degrees of freedom are unaffected by the soft emission. 

In a simple example with a purely diagonal reduced density matrix, it is straightforward to compute the entanglement entropy of the soft emitted bosons. The total hard $+$ soft system is in a bipartite pure state, with the partition being between the hard particles and soft bosons, so the entanglement entropy of the bosons is the same as that of the hard particles. Following the calculation in \cite{Park:2014hya,Carney:2016tcs,Grignani:2016igg}, we can simply write down the entropy:
\be
\label{entropyanswer}
S = \sum_{\beta} \left| S^{\Lambda}_{\beta\alpha} \right|^2 \mathcal{F}_{\beta\alpha}  \ln \left[ \left| S^{\Lambda}_{\beta\alpha} \right|^2 \mathcal{F}_{\beta\alpha} \right].
\ee
This sum is infrared-finite; again, $\mathcal{F}$ is given in \eqref{calF}, and the superscript $\Lambda$ means the naive $S$-matrix computed with virtual bosons only of energies greater than $\Lambda$. Given the explicit form of $\mathcal{F}$, we see that the entropy scales like the log of the infrared detector resolution $E$.

\emph{Discussion.} According to the solution of the infrared catastrophe advocated in \cite{Bloch:1937pw,suura,Weinberg:1965nx}, an infinite number of very low-energy photons and gravitons are produced during scattering events. We have shown that if taken seriously, considering this radiation as lost to the environment completely decoheres almost any momentum state of the outgoing hard particles. The basic idea is simple: the radiation is essentially classical, so any two scattering events are easy to distinguish by their radiation.

The physical content of this result is somewhat unclear. A conservative view is that the methodology of \cite{Bloch:1937pw,suura,Weinberg:1965nx} is ill-suited to finding outgoing density matrices. As remarked earlier, in this formalism, one \emph{must} trace the radiation to get well-defined transition probabilities. An alternative would be to use the infrared-finite $S$-matrix program \cite{Chung:1965zza,kibble,Kulish:1970ut,Ware:2013zja}, in which no trace over radiation is needed at all. But then we need to understand where the physical low-energy radiation is within that formalism--since after all, a photon that is lost to the environment certainly does decohere the system.

The decoherence found here is for the momentum states of the particles: at lowest order in their momenta, soft bosons do not lead to decoherence of spin degrees of freedom. However, the sub-leading soft theorems \cite{Low:1954kd,GellMann:1954kc,Lysov:2014csa} do involve the spin of the hard particles, so going to the next order in the soft particles would be interesting.\footnote{We understand that Strominger has confirmed this. (Private communication)} We would also like to understand to what extent our answers depend on the infinite-time approximation used in the $S$-matrix approach.

To end, we comment on potential applications to the black hole information paradox. The idea advocated in \cite{Hawking:2016msc,Hawking:2016sgy} is that correlations between the hard and soft particles mean that information about the black hole state can be encoded into soft radiation. In \cite{Mirbabayi:2016axw,Gabai:2016kuf,Bousso:2017dny}, the dressed-state formalism and soft factorization has been used to argue that the soft particles simply factor out of the $S$-matrix and thus contain no such information. In the approach used here, it is manifest that the outgoing hard state and outgoing soft state are highly correlated, leading to the decoherence of the hard state. The outgoing density matrix for the hard particles, while not completely thermal, has been mixed in momentum as much as possible while retaining consistency with standard QED/perturbative gravity predictions. It is tempting to conjecture that this generalizes to all asymptotically measurable quantum numbers.

At high center-of-mass energies $\sqrt{s}$, black holes should have production cross-sections given by their geometric areas $\sigma_{prod} \sim \pi r_h^2(\sqrt{s})$.\cite{Banks:1999gd} Using this in \eqref{entropyanswer}, one obtains a hard-soft entanglement entropy scaling like the black hole area times logarithmic soft factors. In this sense one might view the soft radiation as containing a significant fraction of the black hole entropy.

\emph{Acknowledgements.} We thank Scott Aaronson, Tim Cox, Colby Delisle, Bart Horn, Raphael Flauger, John Preskill, Philip Stamp, Bill Unruh, and Jordan Wilson for discussions. After the first version of this paper was submitted to the arXiv, we benefited from correspondence with Andy Strominger and M.M. Sheikh-Jabbari. All of us are grateful for support from NSERC, and DC from the Templeton Foundation and the Pacific Institute of Theoretical Physics.

\appendix

\textbf{Appendix.} Here, we show that the exponents $\Delta A, \Delta B$ controlling the infrared divergences are always positive or zero, and give necessary and sufficient conditions for these exponents to vanish.


The first step is to notice that the expressions for the differential exponents \eqref{differentialexps} between the processes $\alpha \to \beta$ and $\alpha \to \beta'$ are the same as the exponents \eqref{ABdef} for the divergences in the process $\beta \to \beta'$, that is
\begin{align}
\begin{split}
\label{DeltaAstatement}
\Delta A_{\beta\beta',\alpha} & = A_{\beta',\beta}/2, \\
\Delta B_{\beta\beta',\alpha} & = B_{\beta',\beta}/2.
\end{split}
\end{align}
To see this, note from the definitions \eqref{tildeABdef},\eqref{ABdef}, and \eqref{differentialexps} that there are terms in each of $A_{\beta,\alpha}, A_{\beta',\alpha}$, and $\tilde{A}_{\beta\beta',\alpha}$ coming from contractions between pairs of incoming legs, pairs of an incoming and outgoing leg, and pairs of outgoing legs. One can easily check that the in/in and in/out terms cancel pairwise between the $A$ and $\tilde{A}$ terms in $\Delta A$. The remainder is the terms involving contractions between pairs of outgoing legs:
\be
\label{deltaA1}
\Delta A_{\beta\beta',\alpha} = \frac{1}{2} \sum_{p,p' \in \beta} \gamma_{pp'} + \frac{1}{2} \sum_{p,p' \in \beta'} \gamma_{pp'} - \sum_{p \in \beta, p' \in \beta} \gamma_{pp'}
\ee
where we defined $\gamma_{pp'} = e_p e_{p'} \beta_{pp'}^{-1} \ln [(1+\beta_{pp'})/(1-\beta_{pp'})]$. We have used the fact that every $\eta_p$ that would have been in \eqref{deltaA1} is a $-1$ since every line being summed is an outgoing particle, cf. \eqref{softsumresult}. But then we have a relative minus sign and factor of $2$ between the first two terms and the third; this is precisely the same factor that would have come from the relative $\eta_{in} = -1$ and $\eta_{out} = +1$ terms in exponent for the process $\beta \to \beta'$, namely
\be
A_{\beta',\beta} = \sum_{p,p' \in \beta} \gamma_{pp'} + \sum_{p,p' \in \beta'} \gamma_{pp'} -2 \sum_{p \in \beta, p' \in \beta'} \gamma_{p p'}.
\ee
This proves \eqref{DeltaAstatement} for $\Delta A$; an identical combinatorial argument shows that the gravitational exponent obeys the analogous relation, $\Delta B_{\beta\beta',\alpha} = B_{\beta',\beta}/2$. 

Now we prove that for the process $\alpha \to \beta + (soft)$ the exponent $A_{\beta\alpha}$ is always greater or equal to zero with equality if and only if the in and outgoing currents agree; we can then take $\alpha = \beta'$ to get the results quoted in the text. Referring to Weinberg's derivation \cite{Weinberg:1965nx}, we can write $A_{\beta\alpha}$ as
\begin{align}
A_{\beta\alpha} = \frac{1}{2(2\pi)^3} \int_{S^2} d\hat{q} \ t^{\mu}(\hat{q}) t_{\mu}(\hat{q}).
\end{align}
Here,
\begin{align}
\label{eq:currentDecomposition}
t^\mu(\hat{q}) \equiv \sum_n \frac{e_n \eta_n p_n^\mu}{p_n \cdot q} = c(q) q^\mu + c_i(q) (q^i_\perp)^\mu.
\end{align}
In this equation, we have defined a lightlike vector $q^{\mu} = (1,\hat{q})$ and $q^i_\perp$, $i = 1,2$ are two unit normalized, mutually orthogonal, purely spatial vectors perpendicular to $q^\mu$. The sum on $n \in \alpha,\beta$ runs over in- and out-going particles. By charge conservation, $t \cdot q = 0$, which justifies the decomposition in the second equality in \eqref{eq:currentDecomposition}. With this decomposition we may write
\begin{align}
A_{\beta\alpha} = \frac{1}{2(2\pi)^3} \int_{S^2} d\hat q ( c_1^2(q) + c_2^2(q)) \geq 0,
\end{align}
which immediately proves the statement that $A_{\beta\alpha } \geq 0$.

Now it remains to be shown that equality holds if and only if all of the in- and out-going currents match. From the previous paragraph we know that $A_{\beta\alpha}$ vanishes if and only if both $c_i(q) = 0$ for all $q$, that is if and only if $t \cdot q_\perp^i = 0$. Assume that $A_{\beta\alpha} = 0$, so that $q_{\perp} \cdot t(q) = 0$. Now suppose also that $j_{\mb{v}_0}(\alpha) \neq j_{\mb{v}_0}(\beta)$ for some $\mb{v}_0$, where these are the eigenvalues of $j_{\mb{v}} \ket{\alpha} = j_{\mb{v}}(\alpha) \ket{\alpha}$ and similarly for $\beta$. We derive a contradiction. For any finite set of velocities, the functions $f_{\mb{v}}(\hat{\mb{q}}) = (\mb{v}\cdot \mb{q}_{\perp})/(1-\mb{v} \cdot \hat{\mb{q}})$ are linearly independent. Therefore the terms in
\be
0 = t \cdot q_{\perp} = \sum_n \frac{e_n \eta_n v_n \cdot q_{\perp}}{v_n \cdot q}
\ee
must cancel separately for each velocity in the list of $\mb{v}_n$. Consider in particular the term for $\mb{v}_0$. For this to vanish, the sum of the coefficients must vanish, i.e.
\be
0 = \sum_{n | v_{n} = v_0} e_n \eta_n = \left[ j_{v_0}(\alpha) - j_{v_0}(\beta) \right],
\ee
the relative minus coming from the $\eta$ factors. But this contradicts our assumption that $j_{\mb{v}_0}(\alpha) \neq j_{\mb{v}_0}(\beta)$. This completes the proof for $A$.

The proof for gravitons goes similarly. Again referring to Weinberg we write $B$ as
\begin{align}
\label{eq:B}
B_{\beta\alpha} = \frac{G}{4 \pi^2} \int_{S^2} d\hat q t^{\mu\nu} D_{\mu\nu\rho\sigma}t^{\rho\sigma}.
\end{align}
Here, $D_{\mu\nu\rho\sigma} = \eta_{\mu\nu} \eta_{\rho\sigma} - \eta_{\mu\rho} \eta_{\nu\sigma}- \eta_{\mu\sigma} \eta_{\nu\rho}$ is the numerator of the graviton propagator, and
\begin{align}
t^{\mu\nu} = \sum_n \frac{\eta_n p_n^\mu p_n^\nu}{p_n\cdot q} = c q^{(\mu} q^{\nu)} + c^i q^{(\mu} q_{\perp, i}^{\nu)} + c^{ij}q_{\perp, i}^{(\mu}q_{\perp, j}^{\nu)}.
\end{align}
This symmetric tensor obeys $t^{\mu\nu} q_\nu = 0$ by energy-momentum conservation, which justifies the decomposition in the second equality. Using this we have 
\begin{align}
 t^{\mu\nu} D_{\mu\nu\rho\sigma}t^{\rho\sigma} = 2 c^i_j c^j_i - \left( c^i_i \right)^2 = (\lambda_1 - \lambda_2)^2
\end{align}
with $\lambda_{1,2}$ the two eigenvalues of the matrix $c^{ij}$. Plugging this into \eqref{eq:B} we immediately see that $B \geq 0$. The condition for vanishing of $B_{\beta\beta'}$ is that the eigenvalues are equal $\lambda_1 = \lambda_2$, which means that $c^{ij}$ is proportional to the identity matrix. Hence, if $B$ vanishes we have that
\begin{align}
0 = t^{\mu\nu} q^{\perp,1}_\mu q^{\perp,2}_\nu = \sum_n \eta_n E_n \frac{(v_n \cdot q_{\perp}^1) (v_n \cdot q_{\perp}^2)}{v_n \cdot q}. 
\end{align}
As before, any finite set of functions $g_v(q) = (v \cdot q_{\perp}^1) (v \cdot q_{\perp}^2)/(v \cdot q)$ are linearly independent functions of $q$, and so by direct analogy with the previous proof, $B = 0$ if and only if $j^{grav}_{\mb{v}}(\alpha) = j^{grav}_{\mb{v}}(\beta)$ for every $\mb{v}$.


\bibliography{iqi-references}

\end{document}